\documentclass[prb,nofootinbib,twocolumn,superscriptaddress]{revtex4} 

\usepackage{graphicx}
\usepackage{dcolumn}
\usepackage{bm}
\usepackage{threeparttable}
\usepackage{times}
\usepackage{mathptmx}
\usepackage{lscape}
\usepackage{natbib}
\usepackage{amsmath}
\usepackage{amssymb}
\usepackage{braket}
\usepackage{comment}
\usepackage{color}

\def\degree{\kern-.2em\r{}\kern-.3em}

\begin{document}


\title{ Trends in Canonical Nonlinearity for FCC-based Equiatomic Binary Alloys }

\author{Koretaka Yuge}
\affiliation{
Department of Materials Science and Engineering,  Kyoto University, Sakyo, Kyoto 606-8501, Japan\\
}%

\author{Ikumi Nishihara}
\affiliation{
Department of Materials Science and Engineering,  Kyoto University, Sakyo, Kyoto 606-8501, Japan\\
}%

\begin{abstract}
{  
For classical discrete systems under constant composition (typically reffered to as substitutional alloys), canonical average $\phi$ can act as a map from a set of many-body interatomic interactions to that of configuration in thermodynamic equilibrium, the so-called ``canonical nonlinearity: CN'', which generally exhibits complicated nonlinearity. 
Whereas our recent study reveals that the CN can be reasonablly addressed for individual microscopic configuration by two different measures of special vector field on configuration space\cite{asdf1, asdf2} and Kullback-Leibler (KL) divergence $D_{\textrm{KL}}$,\cite{dkl} their direct correlation on real lattices, is still unclear. 
We here address this problem for fcc-based equiatomic binary alloys that have been one of the most studied system in the context of CN. 
We confirm that while local contribution to CN from $D_{\textrm{KL}}$ for each configuration exhibits strong, positive correlation with the vector field, non-local contribution from $D_{\textrm{KL}}$ exhibit no effective correlation. 
We find that depedence of the averaged non-local nonlinearity over all configurations can be well-characterized by normalied geometric distance in configurational polyhedra between for practical and separable system in terms of the structural degrees of freedom. This fact certainly indicates that non-local nonlinearity has profound connection to the geometric configuration for ground-state structures of alloys on configuration space. 
}
\end{abstract}


\maketitle

\section{Introduction}
When we consider substitutional alloys as classical discrete systems under constant composition, microscopic configuration along chosen coordination $Q_{p}$ in thermodynamic equilibrium can be typically given by the canonical average:
\begin{eqnarray}
\Braket{Q_{p}}_{Z} = Z^{-1} \sum_{i} Q_{p}^{\left( i \right)} \exp\left( -\beta U^{\left( i \right)} \right),
\end{eqnarray}
where $Z$ denotes partition function, $\beta$ inverse temperature, $U$ potential energy and summation is taken over all possible configurations. 
For alloys, $U$ can be exactly expressed as the appropriate complete orthonormal basis such as generalized Ising model (GIM),\cite{ce} namely, 
\begin{eqnarray}
\label{eq:gim}
U^{\left( k \right)} = \sum_{j} \Braket{U|Q_{j}} Q_{j}^{\left( k \right)},
\end{eqnarray}
where $\Braket{\cdot|\cdot}$ denotes inner product, i.e., trace over possible configurations. 
Eq.~\eqref{eq:gim} naturally provides the concept that canonical average $\phi$ acts as a map from a set of potential energy $\mathbf{U}$ to equilibrium configuration $\mathbf{Q_{Z}}$:
\begin{eqnarray}
\phi\left( \beta \right) : \mathbf{U} \mapsto \mathbf{Q_{Z}},
\end{eqnarray}
which generally exhibits complicated nonlinearity (hereinafter we call ``canonical nonlinearity (CN)''). 

To multilateraly address the CN, we have introduced two concepts of ``anharmonicity in structural degree of freedoms (ASDF)'' that is a special vector field on configuration space, and Kullback-Leibler divergence $D_{\textrm{KL}}$ on statistical manifold, which is the extention of ASDF to include further non-local CN information. We also confirm that for binary system, the latter one can be further decomposed into two contributions, i.e., local contribution of $D_{\textrm{KL}}^{\textrm{dG}}$ and non-local contribution of$D_{\textrm{KL}}^{\textrm{ns}}$. 
While we recently bridge the above two concepts of CN on configuration space and statistical manifold through stochastic thermodynamics, their direct correlation on real systems is still totally unclear. 

We here address this problem, to investigate how CN as vector field on configuration space and as KL divergence on statistical manifold correlates, and how their correlations are dominated, on fcc-based equiatomic binary alloys that have been most amply studied in the context of CN. 
We confirm that while local contribution in statistical manifold, $D_{\textrm{KL}}^{\textrm{dG}}$, exhibits significant positive correlation with the vector field, non-local contribution of $D_{\textrm{KL}}^{\textrm{ns}}$ does not exhibit effective correlation. 
We find that average of the non-local contribution over possible configurations show clear positive correlation with geometric  distance in configurational polyhedra between practical and SDF-separable system. The details are shown below.

\section{Concepts and Discussions}
\subsection{Brief Concepts for Canonical Nonliearity}
Before we provide basic concepts for the CN, we first briefly explain the GIM that is employed throughout the paper. 
We here focus on a A-B binary system, where the occupation of lattice site $i$ by A (B) is given by the spin variable $\sigma=+1$ ($-1$).
Then information about any given microscopic configuration $k$ along chosen coordination $j$ can be given by
\begin{eqnarray}
\label{eq:gb}
Q_{j}^{\left( k \right)} = \Braket{ \prod_{i\in S_{j}} \sigma_{i} }_{k},
\end{eqnarray}
where the product is performed over lattice points in figure $j$, and $\Braket{\cdot}_{k}$ denotes taking linear average over symmetry-equivalent figures to $j$ in configuraion $k$: Eq.~\eqref{eq:gb} form complete orthonormal basis functions, providing exact expansion of potential energy as given in Eq.~\eqref{eq:gim}. 

Using the GIM basis, we can introduce the measure of CN in terms of the following vector field, ASDF, on configuration space:
\begin{eqnarray}
\label{eq:a}
\mathbf{A}\left( \mathbf{Q} \right) = \left\{ \phi\left( \beta \right) \circ \left( -\beta\Gamma \right)^{-1} \right\}\cdot \mathbf{Q} - \mathbf{Q},
\end{eqnarray}
where $\Gamma$ denotes covariance matrix for configurational density of states (CDOS) before applying many-body interaction to the system. 
The ASDF has significant features: (i) It is independent of energy and temperature, and (ii) it exhibits zero vector when $\phi$ is locally  linear map at $\mathbf{Q}$. Therefore, ASDF is a natural measure of the CN depending only on geometric information derived from the underlying lattice. 

Next, we introduce another measure of the CN on statistical manifold , which is the natural, conceptual extention of ASDF including futher non-local information. We have shown that the following KL divergence corresponds to the extention for CN for binary systems:
\begin{eqnarray}
D_{\textrm{KL}}\left( g_{C}^{\mathbf{Q}} : g_{L}^{\mathbf{Q}} \right) = D_{\textrm{KL}}^{\textrm{dG}}\left( g_{C0}^{\mathbf{Q}} : g_{L}^{\mathbf{Q}} \right) + D_{\textrm{KL}}^{\textrm{ns}}\left( g_{C}^{\mathbf{Q}} : g_{C0}^{\mathbf{Q}} \right)
 \end{eqnarray}
where the first and the second terms of the r.h.s. respectively corresponds to local and non-local contribution to the CN around $\mathbf{Q}$: We just write $D_{\textrm{KL}}^{\rm{dG}}$ and $D_{\textrm{KL}}^{\textrm{ns}}$ hereinafter. 
$g_{C}^{\mathbf{Q}}$, $g_{L}^{\mathbf{Q}}$ and $g_{C0}^{\mathbf{Q}}$ respectively denotes canonical distribution for practical system derived from configuration $\mathbf{Q}$, i.e., $\left\{ \phi\left( \beta \right)\circ \left( -\beta\Gamma \right)^{-1} \right\}\cdot\mathbf{Q}$, that for linear system whose CDOS takes discrete Gaussian with covariance matrix of $\Gamma$ same as the practical system, and the product of marginal distributions for $g_{C}^{\textrm{Q}}$. 
We emphasize that the local contribution to CN explicitly depends on ASDF while the non-local contribution is independent of ASDF. 

Here we focus on the correlation between ASDF and CN as KL divergence for fcc-based equiatomic binary alloys with five pairs of pair correlations that have been most amply studied in the context of CN: 1st-nearest neighbor (1NN)-2NN, 1NN-3NN, 1NN-4NN, 2NN-8NN and 3NN-5NN pairs. For numerical calculations, we prepare 864-atom fcc-based supercell (i.e., $6\times 6\times 6$ expansion of conventional 4-atom cell), that is applied to MC simulation to obtain canonical distribution for individual configuration $\mathbf{Q}$ based on Eq.~\eqref{eq:a} to estimate ASDF and KL divergences. 

\begin{figure}[h]
\begin{center}
\includegraphics[width=0.9\linewidth]{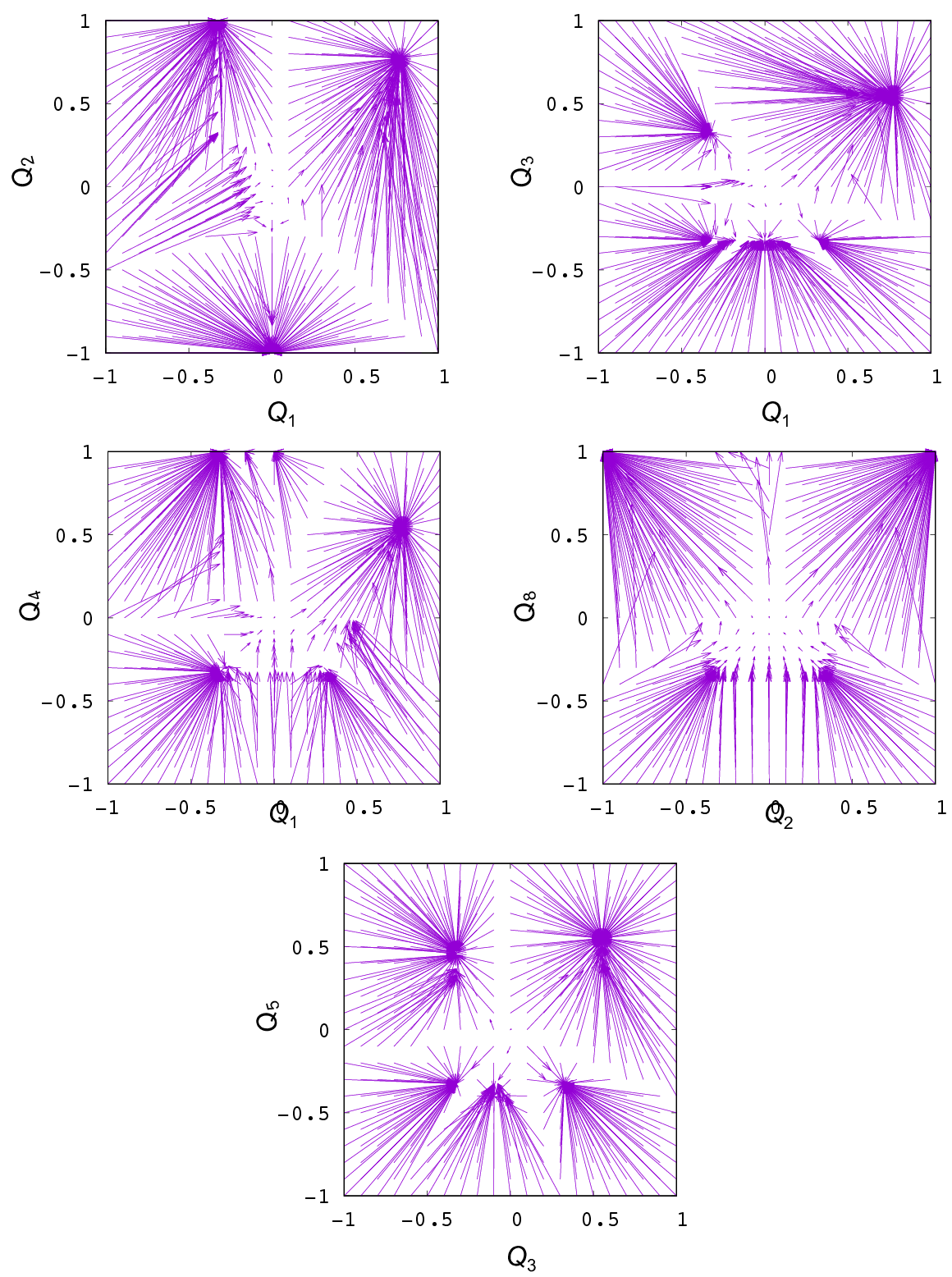}
\caption{ ASDF vector field on configuration space for pair correlations on fcc binary alloys.  }
\label{fig:asdf}
\end{center}
\end{figure}

\begin{figure}[h]
\begin{center}
\includegraphics[width=0.95\linewidth]{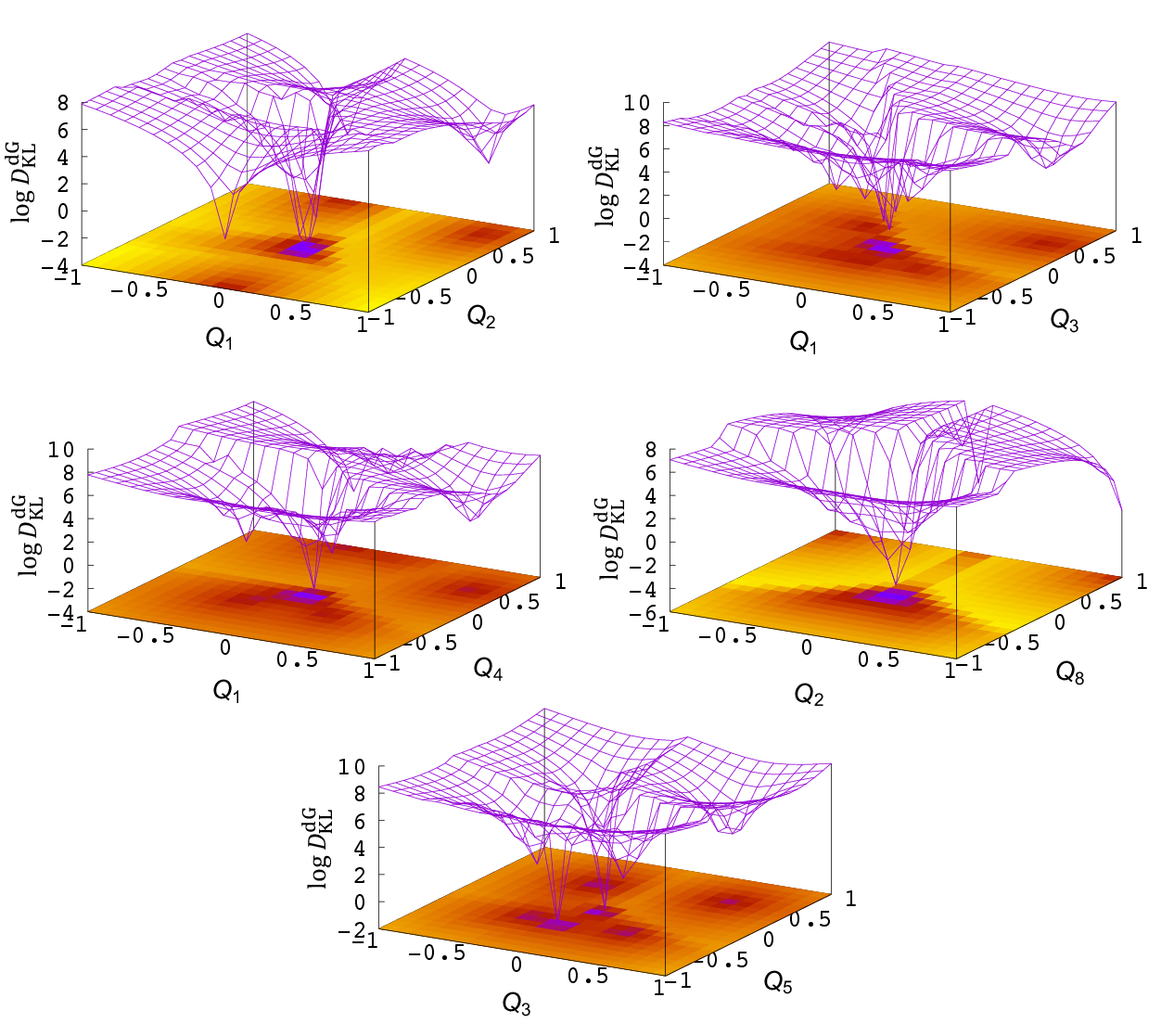}
\caption{ Log plot of contribution to CN from deviation in CDOS from Gaussian, $D_{\textrm{KL}}^{\textrm{dG}}$. }
\label{fig:dG}
\end{center}
\end{figure}

\begin{figure}[h]
\begin{center}
\includegraphics[width=0.95\linewidth]{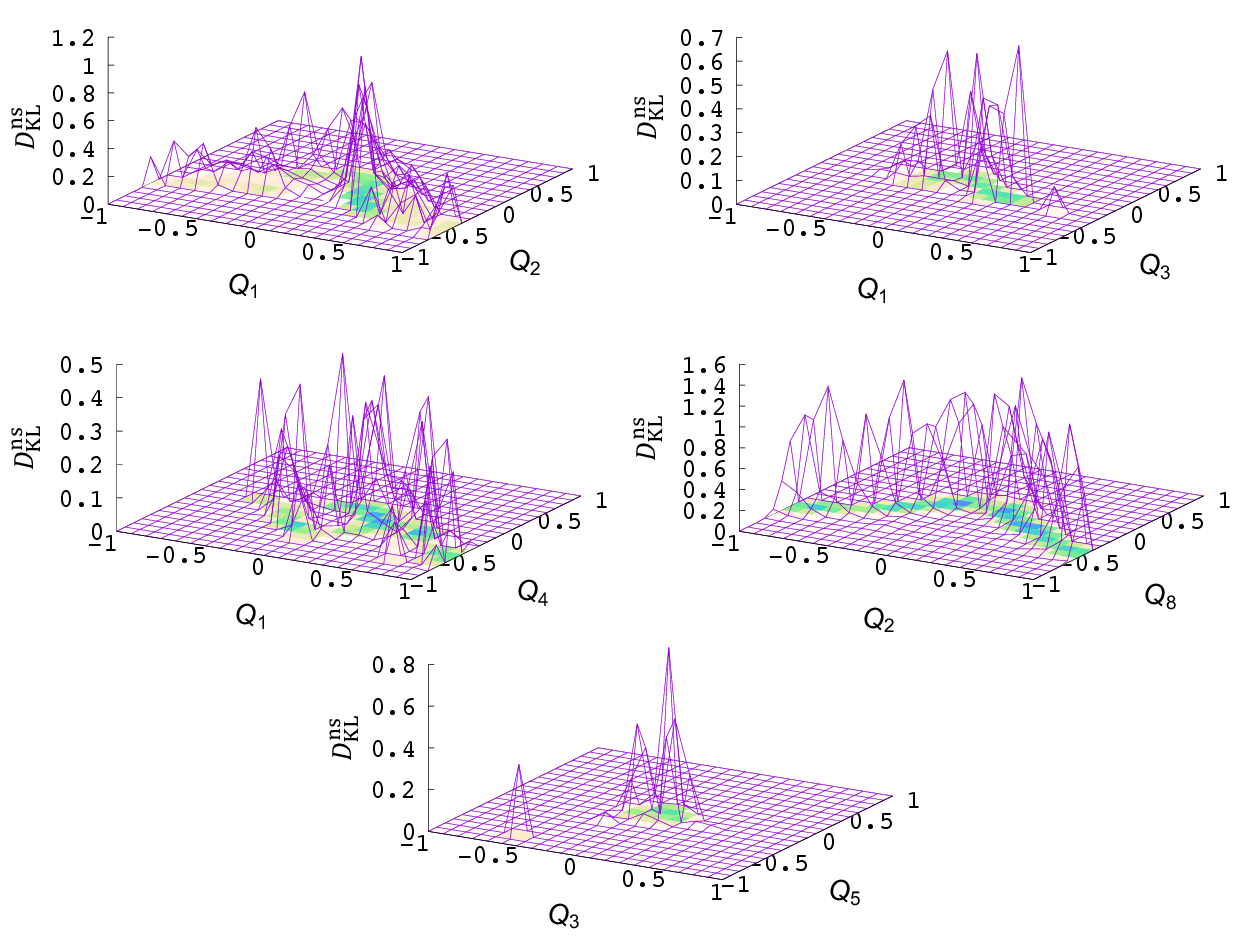}
\caption{ Contribution to CN from nonseparability in SDF, $D_{\textrm{KL}}^{\textrm{ns}}$.}
\label{fig:ns}
\end{center}
\end{figure}

\subsection{Results and Discussions}
\subsubsection{Overall behavioer of ASDF and KL divergence}
We first show in Fig.~\ref{fig:asdf} the overall behavior of ASDF for the five sets of pairs in individual configuraion space. 
Near origin (i.e., random configurationa), absolute of ASDF tends to exhibit smaller value than outer region, naturally reflecting that $\phi$ locally acts as linear map around the random configuration.\cite{bd} From the figure, we can see several points that practically act as adsorption points, typically corresponding to the vetices of configurational polyhedra (i.e., convex polyhedron deriving from end points of ASDF within the prepared area, corresponding to the candidate of ground-state configurations), which has also been seen for other binary systems in our previous studies.\cite{bd} 
Meanwhile, far from the origin, ASDF basically tends to end up to one of the absorption points. 
Namely, we see that the behavior of the ASDF can be qualitatively divided into three characteristic areas, i.e., (i) around the origin (random configuration), almost zero-vectors appears due mainly to the linear character of $\phi$ at the origin, (ii) around the vertices of configurational polyhedra (CP), which is candidate of ground-state configurations, they act as adsorption points, and (iii) around intermediately-ordered configurations, their vectors typically end up around the vertices of the CP.  

Next, we show results of CN as KL divergence in Figs.~\ref{fig:dG} and~\ref{fig:ns} respectively corresponds to 
$D_{\textrm{KL}}^{\textrm{dG}}$ and $D_{\textrm{KL}}^{\textrm{ns}}$. 
For the former of local contribution, around near origin, it takes extremely smaller value than outer region, while around adsorption points of ASDF, it tends to exhibit local minima, and around intermediately-ordered configuration, it takes larger value than others. These appears qualitatively similar tendency to ASDF, which can be naturally accepted since the local contribution of $D_{\textrm{KL}}^{\textrm{dG}}$ is a straightforward extention of ASDF concepts to statistical manifold.\cite{dkl}  

However, for $D_{\textrm{KL}}^{\textrm{ns}}$, its behavior appears totally different from $D_{\textrm{KL}}^{\textrm{dG}}$: i.e., it exhibits sharp local maxima around specific configurations, where their localtion strongly depends on the set of pair correlations. 
Actually, such behaviors of the local contribution of $D_{\textrm{KL}}^{\textrm{ns}}$ do not appear clear correlation with ASDF, which is further discussed later.

\subsubsection{Correlation in CN as between vector field and KL divergence}
\begin{figure}[h]
\begin{center}
\includegraphics[width=1.02\linewidth]{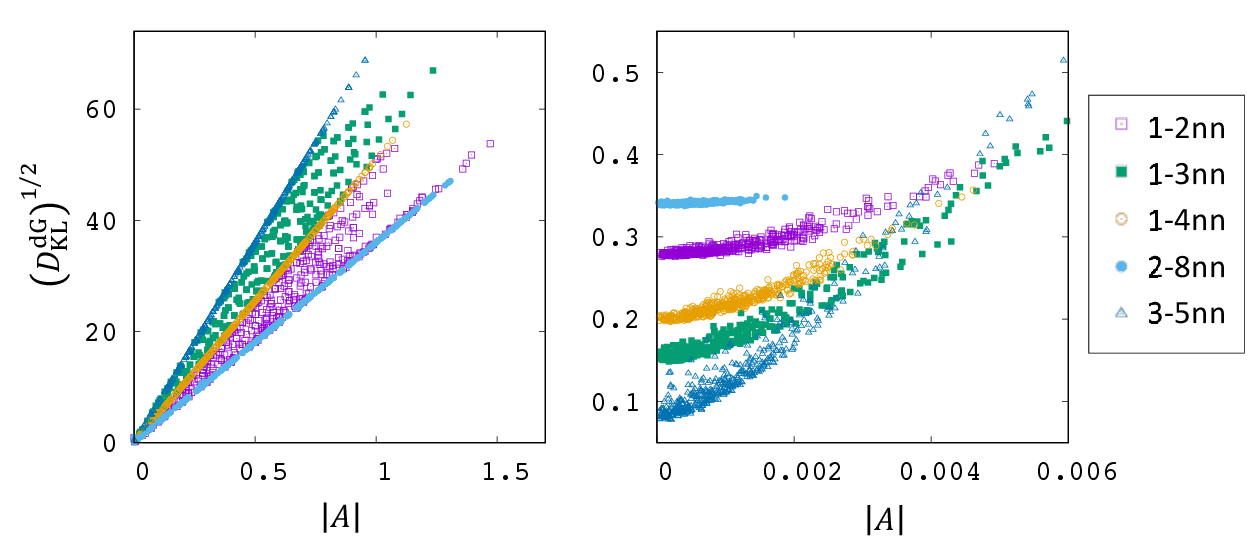}
\caption{ Square root of $D_{\textrm{KL}}^{\textrm{dG}}$ as a function of the absolute of ASDF on each configuration for overall range (left) and near disordered state (right).  }
\label{fig:dga}
\end{center}
\end{figure}

\begin{figure}[h]
\begin{center}
\includegraphics[width=0.77\linewidth]{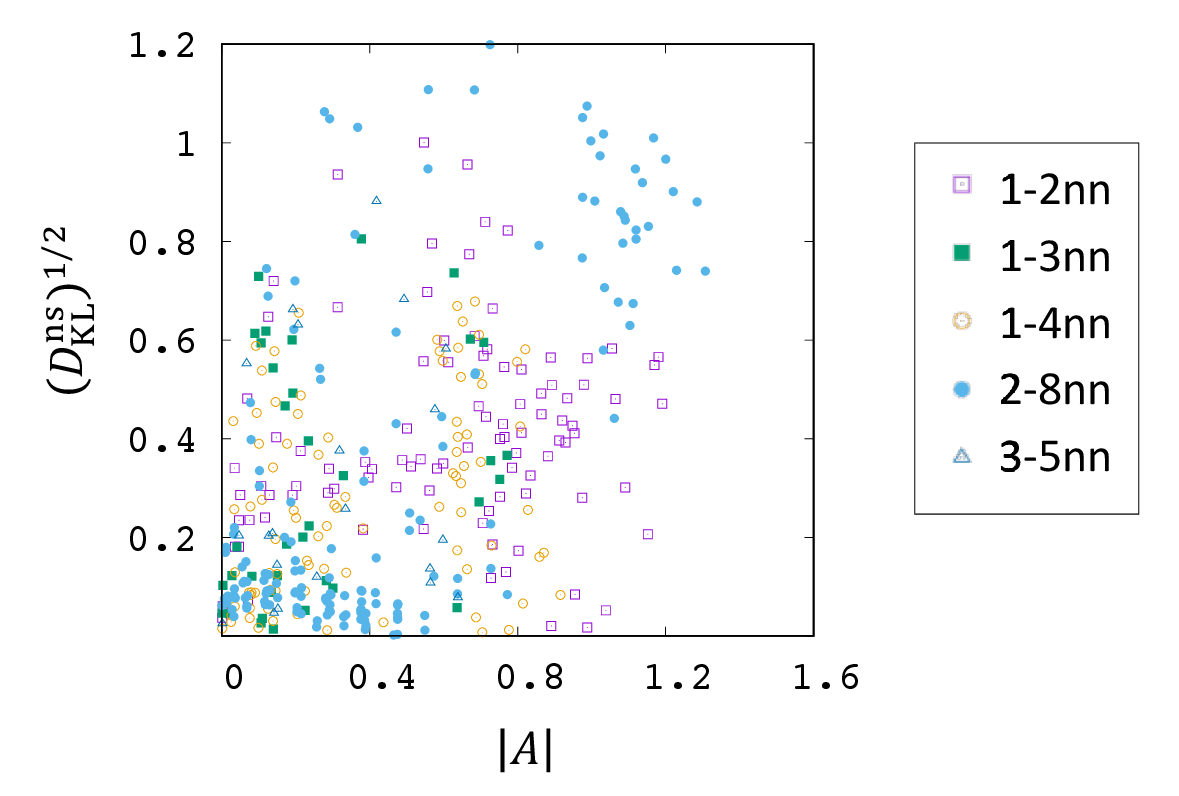}
\caption{Square root of $D_{\textrm{KL}}^{\textrm{ns}}$ as a function of the absolute of ASDF on each configuration.}
\label{fig:nsa}
\end{center}
\end{figure}

We show in Fig.~\ref{fig:dga} correlation between $D_{\textrm{KL}}^{\textrm{dG}}$ and ASDF on each configuration for overall region (left)  and near origin (right) of configuration space. 
These figures indicate that the contribution of $D_{\textrm{KL}}^{\textrm{dG}}$ clearly exhibits positive correlation to ASDF, which is again naturally accpeted since the $D_{\textrm{KL}}^{\textrm{dG}}$ includes straightforwardly extended information about ASDF, i.e., it mainly reflects local CN information around the given configuration similar to the ASDF. 
Meanwhile, the correlation exhibit clear dependence on the set of pairs, which appears to well correlate to the set of coordination number: The gradient of the correlation in descending order is 3-5NN, 1-3NN, 1-4NN, 1-2NN and 2-8NN, and their corresponding coordination numbers for 1, 2, 3, 4, 5 and 8-NN are 12, 6, 24, 12, 24 and 6, respectively.

The correlation between $D_{\textrm{KL}}^{\textrm{dG}}$ and ASDF near origin also appears well correlates to the coordination number, whilst its magnitude is opposite to the overall one.

To further address how the different correlations between $D_{\textrm{KL}}^{\textrm{dG}}$ and ASDF are dominated, we here assume  simple model where canonical distribution for pracitcal and linear systems are both given by normal distributions, and their variance is simply proportional to the corresponding CDOS. Since we measure the divergence on $e$-flat manifold, their canonical distributions are also separable. Therefore, we can straightforwardly rewrite $D_{\textrm{KL}}^{\textrm{dG}}$ as
\begin{widetext}
\begin{eqnarray}
D_{\textrm{KL}}^{\textrm{dG}} \simeq \underbrace{  \frac{A_{x}^{2}}{ 2\sigma_{LX}^{2} } + \frac{A_{y}^{2}}{2\sigma_{LY}^{2}}  }_{f\left( \vec{\sigma} \right)} +   \underbrace{   \ln\left( \frac{\sigma_{LX}\sigma_{LY}}{\sigma_{X}\sigma_{Y} } \right) + \frac{ \sigma_{X}^{2}\sigma_{LY}^{2} + \sigma_{Y}^{2}\sigma_{LX}^{2}  }{2\sigma_{LX}^{2}\sigma_{LY}^{2} } - 1   }_{h\left(\vec{\sigma} \right)},
\end{eqnarray}
\end{widetext}
where $A_{x}$ and $A_{y}$ denotes element of ASDF for individual SDF of $X$ and $Y$, $\sigma_{X}$ and $\sigma_{Y}$, denotes standard deviation (SD) for canonical distribution of practical system, and $\sigma_{LX}$ and $\sigma_{LY}$ that of linear system. 

When contribution from absolute of ASDF is dominant (for overall region), $D_{\textrm{KL}}^{\textrm{dG}}\simeq f\left( \vec{\sigma} \right)$. 
At equicomposition, since SD of CDOS is proportional to $J^{-1/2}$ where $J$ denotes coordination number, we get
\begin{eqnarray}
D_{\textrm{KL}}^{\textrm{dG}} \propto \frac{J}{2} \left|A\right|^{2}
\end{eqnarray}
for the case where constituent SDFs has has the same coordination number $J$, which can reasonably capture the characteristic correlation in the l.h.s. of FIg.~\ref{fig:dga}. 

Meanwhile, around origin where abosolute of ASDF can be neglected, we approximate $D_{\textrm{KL}}^{\textrm{dG}}\simeq h\left( \vec{\sigma} \right)$, and consider the condition where SDFs has the same coordination number. In this case, we can rewrite 
\begin{eqnarray}
\label{eq:dln}
\tilde{D}_{\textrm{KL}}^{\textrm{dG}} \simeq \ln X + \frac{1}{X} -1,
\end{eqnarray}
where $X=\sigma_{LX}^{2}/\sigma_{X}^{2}$, and $\tilde{\cdot}$ denotes divergence around the origin. Since (i) variance of canonical distribution for practical system is bounded for CP inscribed to the outer-rectangle CP while that for linear system is not, (ii) Eq.~\eqref{eq:dln} exhibits monotonic increase in $X>1$, and (iii) the bound due to CP is expected to be much more enhanced for lower coordination number system (because of larger variance of CDOS), 
we can deduce that $X$ has larger value for lower-coordination number system around the disordered state, the following relationships can be satisfied:
\begin{eqnarray}
\tilde{D}_{\textrm{KL}}^{\textrm{dG}} \left( Z' \right) < \tilde{D}_{\textrm{KL}}^{\textrm{dG}} \left( Z \right) \quad \left( Z' > Z \right),
\end{eqnarray}
where $Z$ and $Z'$ denotes coordination number: This can capture the opposite correlation between ASDF and DKL in the r.h.s. of Fig.~\ref{fig:dga}. These consideration indicates that DG contains comparable amount of NOL information as ASDF.

\begin{figure}[h]
\begin{center}
\includegraphics[width=0.6\linewidth]{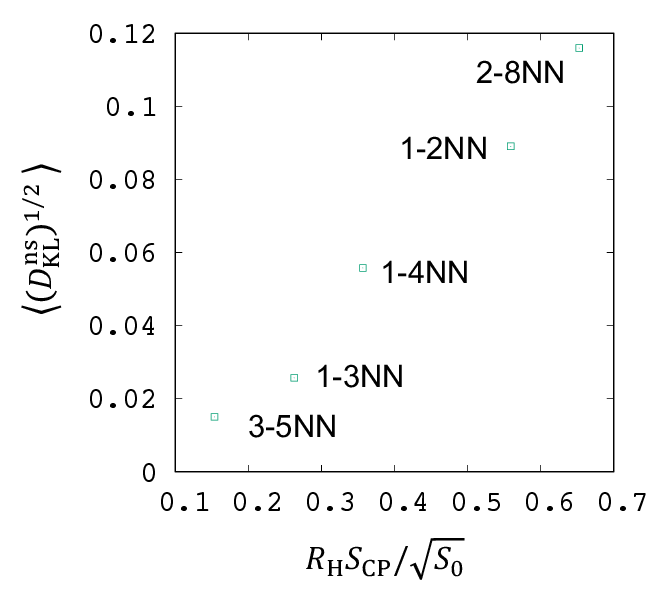}
\caption{ Average of square root of $D_{\textrm{KL}}^{\textrm{ns}}$ over all configuration in terms of the normalized Hausdorff distance between CP and CP0.   }
\label{fig:nsh}
\end{center}
\end{figure}

Next, we discuss about the correlation between ASDF and NS as shown in Fig.~\ref{fig:nsa}. 
As shown in the figure, NS for individual configuration on each system does not appears effective correlation w.r.t. the ASDF, which is naturally accepted since information about the NS is non-local NOL information, and is not explicitly included in the vector field. 
Therefore, we propose alternative strategy to address how the non-local NOL is charaterized by information in configurational geometry. 
Since average of $D_{\textrm{KL}}^{\textrm{ns}}$ over possible configuration under defined configuration space reflects the magnitude of inter-constraints among SDFs, we here focus on the geometric information about the CP for practical system and artificially-constructed separable system: The constraint magnitude on configuration space can be attributed to the difference bewteen that of practical (CP) and separable system (CP0), where we measure the difference by the following asymmetric Hausdorff distance: 
\begin{eqnarray}
\label{eq:hauss}
R_{\textrm{H}} := \max_{a\in \textrm{CP0}} \left\{ \min_{b\in \textrm{CP}} d\left( a,b \right) \right\}.
\end{eqnarray}
Here, $d$ denotes Euclid distance, and the CP0 can be defined by simply taking direct product for configurational polyhedra of constituent SDFs, i.e., the CP0 always takes hyperrectangle (in the present case, rectangle). 
In Eq.~\eqref{eq:hauss}, the reason why we particularly employ asymmetric Hausdorff distance is that CP0 always takes rectangule that takes outer-tangent touch to the practical CP: i.e., we fix the standard of Hausdorff distance always as separable system. 
In order to compare the NS character among different set of SDFs, we would further require additional information for normalizing the Hausdorff distance, i.e., (i) contribution from difference in hypervolumes for separable system, which corresponds to the difference in constraints to individual (separable) SDFs, and (ii) difference in region that can take non-zero probability distribution for $g_{C}$. The former can be regarded as normalizing $R_{\textrm{H}}$ by using the inverse of $\left( V_{\textrm{CP0}} \right)^{1/f}$ ($V_{\textrm{CP0}}$ denotes volume of CP0) with considering the dimension of configuration space, $f$. 
For the latter, when we assume that a set of canonical distributions within each CP have the same amount of information about characterizing corresponding nonlinearity of $D_{\textrm{KL}}^{\textrm{ns}}$, employing the same support for probability distributions in CP0 (i.e., for $g_{C0}$) would lead to requiring normalization coming from the information monotonicity for KL divergence with different coarse-graining: Thus, we here expect that $D_{\textrm{KL}}^{\textrm{ns}}/V_{\textrm{CP}}$ can  reasonablly correlate with the normalized Hausdorff distance, $R_{\textrm{H}}/\left(V_{\textrm{CP0}}\right)^{1/f}$. 
Figure~\ref{fig:nsh} shows the relationship between average of square-root of NS and the normalized Hausdorff distance for sets of SDFs, which exhibits explicit, positive correlations with gradient $M$, namely, 
\begin{eqnarray}
\label{eq:nsh}
\Braket{\left( D_{\textrm{KL}}^{\textrm{ns}} \right)^{1/2}  } \simeq M\cdot\frac{R_{\textrm{H}} V_{\textrm{CP}} }{ \left( V_{\textrm{CP0}} \right)^{1/2} }.
\end{eqnarray}
These results indicate that while non-local information about the nonlinearity, $D_{\textrm{KL}}^{\textrm{ns}}$, does not show effective correlation with the nonlinearity as vector field (reflecting the local information), its average over possible configuration could have profound connection to the shape of configurational polyhedra, corresponding to the geometric configuration of ground-states on configuration space.

\section{Conclusions}
We investigate nonliear character in canonical ensemble of canonical nonlinearity (CN), i.e., the correspondence between a set of potential energy surface and microscopic configuration in thermodynamic equilibrium, based on the correlation between special vector field on configuration space and KL divergences on statistical manifold, which can be decomposed into local and nonlocal information about CN. We find that the local nonlinearity exhibit strong positive correlation with absolute of the vector field, which can be well-characterized in terms of the difference in pair coordination number. Meanwhile, the nonlocal nonlinearity does not exhibit clear correlation to the vector field. The average of nonlocal nonlinearity over all configuration can be well characterized by the normalized Hausdorff distance in configurational polyhedra between practical and SDF-separable system, which indicates that averaged of nonlocal nonlinearity has profound connection to the geometric configuration of ground-state structures in configuration space.

\section{Acknowledgement}
This work was supported by Grant-in-Aids for Scientific Research on Innovative Areas on High Entropy Alloys through the grant number JP18H05453 from the MEXT of Japan and Research Grant from Hitachi Metals$\cdot$Materials Science Foundation.

\end{document}